\documentclass[conference]{IEEEtran}
\usepackage{cite}
\usepackage{graphicx}
\usepackage{psfrag}
\usepackage{subfigure}
\usepackage{url}
\usepackage[cmex10]{amsmath}
\usepackage{array}
\usepackage{graphics}
\usepackage{epstopdf}
\usepackage{amsbsy}
\usepackage{amssymb}
\usepackage{amsthm}

\usepackage{color}
\usepackage{algorithm}
\usepackage{algorithmic}

\begin{document}

\title{On Secure Communication using RF Energy Harvesting Two-Way Untrusted Relay}
%\title{A Tight Radius for Low Complex Sphere Decoding in MIMO Systems using Iterative Matrix Inversion}
%\title{Sphere Decoding in MIMO Systems using Iterative Methods of Matrix Inversion}

\author{\IEEEauthorblockN{Vipul Gupta}
\IEEEauthorblockA{Dept. of EECS\\
University of California, Berkeley, CA, USA\\
E-mail: vipul\_gupta@berkeley.edu}
\and
\IEEEauthorblockN{Sanket S. Kalamkar}
\IEEEauthorblockA{Dept. of EE\\
University of Notre Dame, IN, USA\\
E-mail: skalamka@nd.edu}
\and
\IEEEauthorblockN{Adrish Banerjee}
\IEEEauthorblockA{Dept. of EE\\
IIT Kanpur, India\\
E-mail: adrish@iitk.ac.in}}

\maketitle

\begin{abstract}
We focus on a scenario where two wireless source nodes wish to exchange confidential information via an RF energy harvesting untrusted two-way relay. Despite its cooperation in forwarding the information, the relay is considered untrusted out of the concern that it might attempt to decode the confidential information that is being relayed. To discourage the eavesdropping intention of the relay, we use a friendly jammer. Under the total power constraint, to maximize the sum-secrecy rate, we allocate the power among the sources and the jammer optimally and calculate the optimal power splitting ratio to balance between the energy harvesting and the information processing at the relay. We further examine the effect of imperfect channel state information at both sources on the sum-secrecy rate. Numerical results highlight the role of the jammer in achieving the secure communication under channel estimation errors. We have shown that, as the channel estimation error on any of the channels increases, the power allocated to the jammer decreases to abate the interference caused to the confidential information reception due to the imperfect cancellation of jammer's signal.
\end{abstract}

\section{Introduction}
%The two-way communication was studied for the first time by Shannon in~\cite{shannon}. 
The demand for higher data rates has led to a shift towards higher frequency bands, resulting in higher path loss. Thus relays have become important for reliable long distance wireless transmissions. The two-way relay has received attention in the past few years due to its ability to make communications more spectral efficient~\cite{relay2006,yener_two_way}. In a two-way relay-assisted communication, the relay receives the information from two nodes simultaneously, which it broadcasts in the next slot. 

\subsection{Motivation}

To improve the energy efficiency, harvesting energy from the surrounding environment has become a promising approach, which can prolong the lifetime of energy-constrained nodes and avoid frequent recharging and replacement of batteries. 
%Many natural sources of energy can be used to power such nodes, like solar, thermal, wind etc. However, 
In \cite{lav} and \cite{rui4}, authors have proposed the concept of energy harvesting using radio-frequency (RF) signals that carry information as a viable source of energy. Simultaneous wireless information and power transfer has applications in cooperative relaying. The works in~\cite{nasir,chen1,sanket3,liu1,duong_two_way} study throughput maximization problems when the cooperative relays harvest energy from incoming RF signals to forward the information, where references~\cite{liu1,duong_two_way} have focused on two-way relaying.
% Keeping in mind the surge in demand of energy harvesting nodes as a common source of power for nodes, we investigate the effects of employing energy harvesting techniques in two-way cooperative relay systems in this paper.

Though the open wireless medium has facilitated cooperative relaying, it has also allowed unintended nodes to eavesdrop the communication between two legitimate nodes. Traditional ways to achieve secure communication rely on upper-layer cryptographic methods that involve intensive key distribution. Unlike this technique, the physical layer security aims to achieve secure communication by exploiting the random nature of the wireless channel. In this regard, Wyner introduced the idea of secrecy rate for the wiretap channel, where the secure communication between two nodes was obtained without private keys~\cite{wyner}.
%In~\cite{korner}, authors evaluated the expression for secrecy rate for wireless systems.

For cooperative relaying with energy harvesting, \cite{quan,xing2,chen2016} investigate relay-assisted secure communication in the presence of an external eavesdropper. The security of the confidential message may still be a concern when the source and the destination wish to keep the message secret from the relay, despite its help in forwarding the information~\cite{he,huang,wang,li2,park2015}. Hence the relay is trusted in forwarding the information, but untrusted out of the concern that the relay might attempt to decode the confidential information that is being relayed.\footnote{In this case, the decode-and-forward relay is no longer suitable to forward the confidential information.} In practice, such a scenario may occur in heterogeneous networks, where all nodes do not possess the same right to access the confidential information. For example, if two nodes having the access to confidential information wish to exchange that information but do not have the direct link due to severe fading and shadowing, they might require to take the help from an intermediate node that does not have the privilege to access the confidential information.

\subsection{Related Work}
In~\cite{he}, authors show that the cooperation by an untrusted relay can be beneficial and can achieve higher secrecy rate than just treating the untrusted relay as a pure eavesdropper. In \cite{zhang2012}, authors investigate the secure communication in untrusted two-way relay systems with the help of external friendly jammers and show that, though it is possible to achieve a non-zero secrecy rate without the friendly jammers, the secrecy rate at both sources can effectively be improved with the help from an external friendly jammer. In~\cite{wang2016}, authors have focused on improving the energy efficiency while achieving the minimum secrecy rate for the untrusted two-way relay. The works in~\cite{he,huang,wang,li2,park2015,zhang2012,wang2016} assume that the relay is a conventional node and has a stable power supply.
%In \cite{paper1.3, sanket}, the authors investigate the effects
As to energy harvesting untrusted relaying, the works in~\cite{sanket,meng,mousa} analyze the effect of untrusted energy harvesting one-way relay on the secure communication between two legitimate nodes. To the best of our knowledge, for energy harvesting two-way untrusted relay, the problem of achieving the secure communication has not been yet studied in the literature.  
                                                                                                                                                                                                                                                                                                                                                                                                                         
                                                                                                                                                                                                                                                                                                                                                                                                                         \subsection{Contributions}
\begin{figure}
\centering
\includegraphics[scale=0.271]{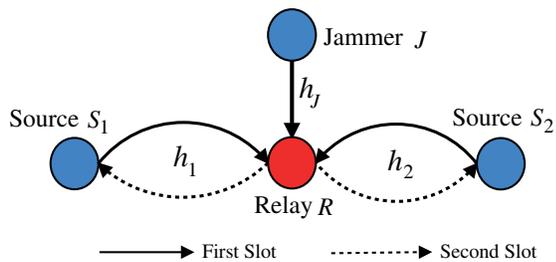}
\caption{\hspace{-1mm}Secure communication via an untrusted energy harvesting two-way relay. $h_{i}$ with $i \in \{1, 2, J\}$ denotes a channel coefficient.}
\label{fig:sys}\vspace*{-2mm}
\end{figure}

The contributions and main results of this paper are as follows: 
\begin{itemize}
\item First, assuming the perfect channel state information (CSI) at source nodes, we extend the notion of secure communication via an untrusted relay for the two-way wireless-powered relay, as shown in Fig.~\ref{fig:sys}. To discourage the eavesdropping intentions of the relay, a friendly jammer sends a jamming signal during relay's reception of signals from source nodes.
\item To harvest energy, the relay uses a part of the received RF signals which consist of two sources' transmissions and the jamming signal. Hence we utilize the jamming signal effectively as a source of extra energy in addition to its original purpose of degrading relay's eavesdropping channel.
\item Under the total power constraint, we exploit the structure of the original optimization problem and make use of the signomial geometric programming technique~\cite{boyd_gp} to jointly find the optimal power splitting ratio for energy harvesting and the optimal power allocation among sources and the jammer that maximize the sum-secrecy rate for two source nodes.
\item Finally, with the imperfect CSI at source nodes, we study the joint effects of the energy harvesting nature of an untrusted relay and channel estimation errors on the sum-secrecy rate and the power allocated to the jammer. We particularly focus on the role of jammer in achieving the secure communication, where we show that the power allocated to the jammer decreases as the estimation error on any of the channels increases, in order to subside the detrimental effects of the imperfect cancellation of the jamming signal at source nodes.
%We analyze the role of the jammer in different conditions and how it effects the total secrecy rate. Our analysis is novel in the sense that the total power utilized in the two-hop transmission is quantified, with several trade-offs being analyzed in a limited power scenario when energy harvesting is employed at the untrusted relay.   

\end{itemize}

%The legitimate nodes operate under the limited power budget for the two-way secure transmission.   

\section{Secure Communication with Perfect CSI}
\subsection{System Model}
Fig.~\ref{fig:sys} shows the communication protocol between two legitimate source nodes $S_1$ and $S_2$---lacking the direct link between them---via an untrusted two-way relay $R$. All nodes are half-duplex and have a single antenna~\cite{zhang2012}. To discourage eavesdropping by the relay, a friendly jammer $J$ sends the jamming signal during relay's reception of sources' signals. The communication of a secret message between $S_1$ and $S_2$ happens over two slots of equal duration $T/2$. In the first slot, the nodes $S_1$ and $S_2$ jointly send their information to the relay with powers $P_1$ and $P_2$, respectively, and the jammer $J$ sends the jamming signal with power $P_J$. The powers $P_1$, $P_2$, and $P_J$ are restricted by the power budget $P$ such that $P_1 + P_2 + P_J \leq P$. This constraint may arise, for instance, when the sources and the jammer belong to the same network, and the network has a limited power budget to cater transmission requirements of sources and the jammer. The relay uses a part of the received power to harvest energy. In the second slot, using the harvested energy, the relay broadcasts the received signal in an amplify-and-forward manner.

Let $h_1$, $h_2$, and $h_J$ denote the channel coefficients of the reciprocal channels from the relay to $S_1$, $S_2$, and jammer $J$, respectively. In this section, we assume that both sources have the perfect CSI for all channels, which can be obtained from the classical channel training, estimation, and feedback from the relay. But if there are errors in the estimation and/or feedback, the sources will have imperfect CSI, which is the focus of Section~\ref{sec:ICSI}. Hence the relay is basically trusted when it comes to providing the services like feeding CSI back to transmitters and forwarding the information but untrusted in the sense that it is not supposed to decode the confidential information that is being relayed~\cite{wang2016}. Both sources have the perfect knowledge of the jamming signal~\cite{zhang2012}.\footnote{Jammer can use pseudo-random codes as the jamming signals that are known to both sources beforehand but not to the untrusted relay.}
% No direct link exists between $S_1$ and $S_2$ they send the data simultaneously. 

\subsection{RF Energy Harvesting at Relay}
The relay is an energy-starved node. It harvests energy from incoming RF signals which include information signals from nodes $S_1$ and $S_2$ and the jamming signal from the jammer. To harvest energy from received RF signals, the relay uses power splitting (PS) policy~\cite{rui4}. In PS policy, the relay uses a fraction $\beta$ of the total received power for energy harvesting.
% In TS policy, the relay uses a fraction of the total time to harvest the energy and processes the information in rest of the time. Our further discussions will be based only on the power splitting technique; but, the results in this paper can be extended for TS policy.
Under PS policy, the energy harvested by the relay is\footnote{For the exposition, we assume that the incident power on the energy harvesting circuitry of the relay is sufficient to activate it.}
\begin{equation}
E_H = \beta \eta \left(P_1|h_1|^2 + P_2|h_2|^2 + P_J|h_J|^2\right)(T/2),
\label{eq:EH}
\end{equation} 
where $\eta$ is the energy conversion efficiency factor with $0 < \eta < 1$. The transmit power of the relay in the second slot is 
\begin{equation}
P_H = \frac{E_H}{T/2} = \beta \eta \left(P_1|h_1|^2 + P_2|h_2|^2 + P_J|h_J|^2\right).
\label{eq:PH}
\end{equation} 

% However, it needs to be allocated power from a limited energy bank, which would lead to a trade-off between secrecy and information throughput (see results in Section \ref{numerical_results}).
 
\subsection{Information Processing and Relaying Protocol}
In the first slot, the relay receives the signal
\begin{equation}
y_R \!=\! \sqrt{(1-\beta)}(\sqrt{P_1}h_1x_1 + \sqrt{P_2}h_2x_2 + \sqrt{P_J}h_Jx_J) + n_R,
\label{eq:rec_rel}
\end{equation}
where $x_1$ and $x_2$ are the messages of $S_1$ and $S_2$, respectively, with $\mathbb{E}[|x_1|^2] = \mathbb{E}[|x_1|^2] = 1$. Also $x_J$ is the artificial noise by the jammer with $\mathbb{E}[|x_J|^2] = 1$, and $n_R$ is the additive white Gaussian noise (AWGN) at relay with mean zero and variance $N_0$. Using the received signal $y_R$, the relay may attempt to decode the confidential messages $x_1$ and $x_2$. To shield the confidential messages $x_1$ and $x_2$ from relay's eavesdropping, we assume that the physical layer security coding like stochastic encoding and nested code structure can be used (see~\cite{wang2016} and \cite{hayashi}). The relay can decode one of the sources' confidential messages, \textit{i.e.}, either $x_1$ or $x_2$, if its rate is such that it can be decoded by considering other source's message as noise~\cite{tekin}. In this case, at relay, the signal-to-noise ratio (SNR) corresponding to $x_1$, \textit{i.e.}, the message intended for $S_2$, is given by
\begin{align}
\text{SNR}_{R_2} = \frac{\widetilde\beta P_1|h_1|^2}{\widetilde\beta P_2|h_2|^2 + \widetilde\beta P_J|h_J|^2 + N_0},
\label{eq:S1}
\end{align}
where $\widetilde\beta = 1-\beta$. Accordingly the achievable throughput of $S_1-R$ link is $C_2^R = (1/2)\log(1 + \text{SNR}_{R_2})$. In \eqref{eq:S1}, the term $\widetilde\beta P_2|h_2|^2$, corresponding to $S_2$'s message for $S_1$ indirectly serves as an artificial noise for the relay in addition to the signal $\widetilde\beta P_J|h_J|^2$ from the jammer. Similarly, the SNR corresponding to $x_2$, \textit{i.e.}, the message intended for $S_1$, is given by
\begin{equation}
\text{SNR}_{R_1} = \frac{\widetilde\beta P_2|h_2|^2}{\widetilde\beta P_1|h_1|^2 + \widetilde\beta P_J|h_J|^2 + N_0},
\end{equation}
where $\widetilde\beta P_1|h_1|^2$ serves as an artificial noise for the relay. Thus the achievable throughput of $S_2-R$ link is $C_1^R = (1/2)\log(1 + \text{SNR}_{R_1})$.
Let $\gamma_i = P_i |h_i|^2/N_0$, where $i \in \lbrace 1,2,J\rbrace$.
It follows that
\begin{equation}\label{snr_relay}
\text{SNR}_{R_2} = \frac{\widetilde\beta\gamma_1}{\widetilde\beta\gamma_2 + \widetilde\beta \gamma_J + 1},~ \text{SNR}_{R_1} = \frac{\widetilde\beta\gamma_2}{\widetilde\beta\gamma_1 + \widetilde\beta \gamma_J + 1}.
\end{equation}
The relay amplifies the received signal $y_R$ given by \eqref{eq:rec_rel} by a factor $\alpha$ based on its harvested power $P_H$. Accordingly,
\begin{align}\label{alpha}
\alpha &= \sqrt{\frac{P_H}{\widetilde\beta P_1|h_1|^2 + \widetilde\beta P_2|h_2|^2 + \widetilde\beta P_J|h_J|^2 + N_0}}\nonumber \\
&= \sqrt{\frac{\beta\eta(\gamma_1 + \gamma_2 + \gamma_J)}{\widetilde\beta\gamma_1 + \widetilde\beta\gamma_2 + \widetilde\beta\gamma_J + 1}}.
\end{align}
The received signal at $S_2$ in the second slot is given by
\begin{equation}
y_2 = h_2(\alpha y_R) + n_2,
\end{equation}
where $n_2$ is AWGN with power $N_0$. We assume that $S_1$ and $S_2$ know $x_J$ beforehand. Hence after cancelling the terms that are known to $S_2$, \textit{i.e.}, the terms corresponding to $x_2$ and $x_J$, the resultant received signal at $S_2$ is
\begin{equation}
y_2 = \underbrace{h_2\alpha\sqrt{\widetilde\beta P_1}h_1x_1}_{\text{desired~signal}} + \underbrace{h_2\alpha n_R + n_2}_{\text{noise}}.
\label{eq:rec_s2}
\end{equation}
The perfect CSI allows $S_2$ to cancel unwanted components of the signal. Substituting $\alpha$ from \eqref{alpha} in \eqref{eq:rec_s2}, we can express the SNR at node $S_2$ as
\begin{equation}
\text{SNR}_{S_2} = \frac{\gamma_1|h_2|^2\beta\widetilde\beta\eta(\gamma_1 + \gamma_2 + \gamma_J)}{(|h_2|^2\beta\eta + \widetilde\beta)(\gamma_1 + \gamma_2 + \gamma_J) + 1},
\end{equation}
and the corresponding achievable throughput of link $R-S_2$ is $C_2^S = (1/2)\log(1 + \text{SNR}_{S_2})$. Similarly the received signal at $S_1$ is
\begin{equation}
y_1 = \underbrace{h_1\alpha\sqrt{\widetilde\beta P_2}h_2x_2}_{\text{desired~signal}} + \underbrace{h_1\alpha n_R + n_1}_{\text{noise}}.
\end{equation} 
The SNR at node $S_1$ is  
\begin{equation}
\text{SNR}_{S_1} = \frac{\gamma_2|h_1|^2\beta\widetilde\beta\eta(\gamma_1 + \gamma_2 + \gamma_J)}{(|h_1|^2\beta\eta + \widetilde\beta)(\gamma_1 + \gamma_2 + \gamma_J) + 1},
\end{equation}
and the corresponding achievable throughput of link $R-S_1$ is $C_1^S = (1/2)\log(1 + \text{SNR}_{S_1})$.

\subsection{Secrecy Rate and Problem Formulation}\label{prob_formulation}
For the communication via two-way untrusted relay, the sum-secrecy rate is given by
\begin{align}\label{secrecy_rate2}
C_S &= \left[C_1^S - C_1^R \right]^+ + \left[C_2^S - C_2^R \right]^+ \nonumber\\
&= \left[\frac{1}{2}\log_2(1+ \text{SNR}_{S_1}) - \frac{1}{2}\log_2(1+ \text{SNR}_{R_1})\right]^+ \nonumber\\
&+ \left[\frac{1}{2}\log_2(1+ \text{SNR}_{S_2}) - \frac{1}{2}\log_2(1+ \text{SNR}_{R_2})\right]^+,
\end{align}
where $[x]^+ \triangleq \max(x,0).$
Given the total power budget $P$, we have a constraint on transmit powers, \textit{i.e.}, $P_1 + P_2 + P_J \leq P$. To maximize the sum-secrecy rate, we optimally allocate powers $P_1$, $P_2$, and $P_J$ to $S_1$, $S_2$, and $J$, respectively, and find the optimal power splitting ratio $\beta$. We can formulate the optimization problem as
\begin{align}
\mathop{\mathrm{maximize}}_{\beta, \widetilde\beta, P_1, P_2,P_J}& ~~~ C_S  \nonumber \\
\mathrm{subject~to} &~~~ P_1 + P_2 + P_J \leq P, \nonumber \\
&~~~ \beta + \widetilde\beta = 1,\nonumber \\
 & ~~~\beta, \widetilde\beta \leq 1, \nonumber \\
  & ~~~\beta, \widetilde\beta, P_1, P_2, P_J  \geq 0.  
\label{eq:opt_prob}
\end{align}\vspace*{-4mm}
%\begin{alignat*}{2}
%\underset{\beta, \widetilde\beta, P_1, P_2,P_J}{\text{maximize }}  & ~~C_S & \text{(Secrecy rate maximization)}  \\
%\text{subject to } & \beta, \widetilde\beta, P_1, P_2, P_J  \geq 0 & \text{(Non-negative power)} \\
%&\beta + \widetilde\beta = 1 & \text{(Definition of $\widetilde\beta$)}\\
%&\beta, \widetilde\beta \leq 1 & \text{(Power splitting ratio $\leq 1$)}\\
%\text{and } & P_1 + P_2 + P_J \leq P & \text{(Limited total power)}.
%\end{alignat*}

\noindent Based on the non-negativeness of two terms in the secrecy rate expression given by \eqref{secrecy_rate2}, we need to investigate four cases. We calculate the sum-secrecy rate in all four cases, with the best case being the one that gives the maximum sum-secrecy rate. 

\subsection*{\underline{Case I}: $C_1^S - C_1^R \geq 0$ and $C_2^S - C_2^R \geq 0$}
Substituting $\gamma_i = P_i|h_i|^2/N_0$ and simplifying the problem in \eqref{eq:opt_prob}, it follows that
\begin{subequations}
\begin{align}
\mathop{\mathrm{minimize}}_{\beta, \widetilde\beta, P_1, P_2,P_J}& ~~~ \frac{1}{2}\log_2\frac{f(\beta,\widetilde\beta, \gamma_1,\gamma_2, \gamma_J)}{g(\beta,\widetilde\beta,\gamma_1,\gamma_2, \gamma_J)} \label{eq:AA} \\
\mathrm{subject~to} &~~~ \frac{\gamma_1N_0}{|h_1|^2} + \frac{\gamma_2N_0}{|h_2|^2} + \frac{\gamma_JN_0}{|h_J|^2} \leq P , \label{eq:BB} \\
&~~~ \beta + \widetilde\beta = 1,\label{eq:CC} \\
 & ~~~\beta, \widetilde\beta \leq 1, \label{eq:DD}\\
  & ~~~\beta, \widetilde\beta, P_1, P_2, P_J  \geq 0\label{eq:EE}, 
\end{align}
\label{eq:opt_prob1}
\end{subequations}\vspace*{-5mm}
%\begin{alignat*}{2}
%\underset{\beta, \widetilde\beta, P_1, P_2,P_J}{\text{minimize }}   & \frac{1}{2}\log\frac{f(\beta,\widetilde\beta, \gamma_1,\gamma_2, \gamma_J)}{g(\beta,\widetilde\beta,\gamma_1,\gamma_2, \gamma_J)}  \\
%\text{subject to } &  \beta, \widetilde\beta, \gamma_1,\gamma_2, \gamma_J  \geq 0\\ 
%&\beta + \widetilde\beta = 1 \\
%& \beta,\widetilde\beta \leq 1\\
%\text{and } & \frac{\gamma_1N_0}{|h_1|^2} + \frac{\gamma_2N_0}{|h_2|^2} + \frac{\gamma_JN_0}{|h_J|^2} \leq P 
%\end{alignat*}

\noindent where
\begin{align*}
f(\beta,\widetilde\beta,\gamma_1,\gamma_2, \gamma_J) &= [\widetilde\beta(\gamma_1 + \gamma_2 + \gamma_J) + 1]^2  \nonumber\\
&\times\lbrack 1 + (\gamma_1 + \gamma_2 + \gamma_J)(|h_2|^2\beta\eta + \widetilde\beta)\rbrack  \nonumber\\
&\times\lbrack 1 + (\gamma_1 + \gamma_2 + \gamma_J)(|h_1|^2\beta\eta + \widetilde\beta)\rbrack,
\end{align*}
\vspace*{-5mm}

\noindent and
\begin{align*}
g(\beta,\widetilde\beta,\gamma_1,\gamma_2, \gamma_J) = (\widetilde\beta(\gamma_2 + \gamma_J) + 1)(\widetilde\beta(\gamma_1 + \gamma_J) + 1) \nonumber\\
\times\lbrack(\gamma_1 + \gamma_2 + \gamma_J)(\widetilde\beta + |h_2|^2\beta\eta(\widetilde\beta\gamma_1 + 1))+1\rbrack\nonumber\\
\times\lbrack(\gamma_1 + \gamma_2 + \gamma_J)(\widetilde\beta + |h_1|^2\beta\eta(\widetilde\beta\gamma_2 + 1))+1\rbrack.
\end{align*}
We can drop the logarithm from the objective \eqref{eq:AA} as it retains the monotonicity and yields the same optimal solution. We introduce an auxiliary variable $t$ and do the following transformation.
\begin{subequations}
\begin{align}
\mathop{\mathrm{minimize}}_{\beta, \widetilde\beta, P_1, P_2,P_J}& ~~~ \frac{f(\beta,\widetilde\beta,\gamma_1,\gamma_2, \gamma_J)}{t} \label{objective} \\
\mathrm{subject~to} &~~~ t \leq g(\beta,\widetilde\beta,\gamma_1,\gamma_2, \gamma_J), \label{g_mono} \\
&~~~ \frac{\gamma_1N_0}{|h_1|^2} + \frac{\gamma_2N_0}{|h_2|^2} + \frac{\gamma_JN_0}{|h_J|^2} \leq P , \label{posy2} \\
&~~~ \beta + \widetilde\beta \leq 1,\label{beta} \\
 & ~~~\beta, \widetilde\beta \leq 1, \label{posy1} \\
  & ~~~t, \beta, \widetilde\beta, P_1, P_2, P_J  \geq 0 \label{chill2}. 
\end{align}
\label{prob}
\end{subequations}
\vspace*{-4mm}

%\begin{subequations}\label{prob}
%\begin{alignat}{3}
%\underset{\beta, \widetilde\beta, P_1, P_2,P_J}{\text{minimize }}   & \frac{f(\beta,\widetilde\beta,\gamma_1,\gamma_2, \gamma_J)}{t}  & \text{($t$ introduced)} \label{objective}\\
%\text{subject to } & t \leq g(\beta,\widetilde\beta,\gamma_1,\gamma_2, \gamma_J)& \text{(Valid transformation)} \label{g_mono}\\
%%& t> 0& \text{(Positive secrecy rate)} \label{chill1}\\
%& \beta,\widetilde\beta,\gamma_1,\gamma_2, \gamma_J, t \geq 0 \label{chill2}\\
%&\beta + \widetilde\beta = 1 \label{beta}\\
%& \beta,\widetilde\beta \leq 1 \label{posy1}\\
%\text{and } & \frac{\gamma_1N_0}{|h_1|^2} + \frac{\gamma_2N_0}{|h_2|^2} + \frac{\gamma_JN_0}{|h_J|^2} \leq P \label{posy2}.
%\end{alignat}
%\end{subequations}
\noindent The above transformation is valid for $t>0$ because, to minimize the objective $f(\beta,\widetilde\beta,\gamma_1,\gamma_2, \gamma_J)/t$, we need to maximize $t$, and it happens when $t = g(\beta,\gamma_1,\gamma_2, \gamma_J)$. Hence under the optimal condition, we have $t = g(\beta,\gamma_1,\gamma_2, \gamma_J)$, and the problems \eqref{eq:opt_prob} and \eqref{prob} are equivalent. Further we can replace the constraint \eqref{eq:CC} by
\begin{equation}\label{beta2}
\beta + \widetilde\beta \leq 1.\vspace*{-1mm}
\end{equation}
The substitution of \eqref{eq:CC} by \eqref{beta2} in problem \eqref{prob} yields an equivalent problem because $\beta + \widetilde\beta = 1$ under the optimal condition, {\em i.e.}, if $\beta + \widetilde\beta < 1$, we can always increase the value of $\beta$ so that $\beta + \widetilde\beta = 1$. The increase in $\beta$ leads to more harvested energy, which in turn increases the transmit power of the relay and the sum-secrecy rate.   

The objective \eqref{objective} is a posynomial function and~\eqref{posy2}, \eqref{beta}, and \eqref{posy1} are posynomial constraints~\cite{boyd_gp}. When the objective and constraints are of posynomial form, the problem can be transformed into a Geometric Programming (GP) form and converted into a convex problem~\cite{boyd_gp}. Also, as the domain of GP problem includes only real positive variables, the constraint \eqref{chill2} is implicit. But the constraint \eqref{g_mono} is not posynomial as it contains a posynomial function $g$ which is bounded from below and GP cannot handle such constraints. We can solve this problem if the right-hand side of \eqref{g_mono}, \textit{i.e.}, $g(\beta,\widetilde\beta,\gamma_1,\gamma_2, \gamma_J)$, can be approximated by a monomial. Then the problem \eqref{prob} reduces to a class of problems that can be solved by Signomial Geometric Programming (SGP)~\cite{boyd_gp}. 

To find a monomial approximation of the form $\widehat g(\mathbf{x}) = c\prod_{i=1}^5x_i^{a_i}$ of a function $g(\mathbf{x})$ where $\mathbf{x} = [\beta,\widetilde\beta,\gamma_1,\gamma_2, \gamma_J]^T$ is the vector containing all variables, it would suffice if we find an affine approximation of $h(\mathbf{y}) = \log g(\mathbf{y})$ with $i$th element of $\mathbf{y}$ given by $y_i = \log x_i$~\cite{boyd_gp}. Let the affine approximation of $h(\mathbf{y})$ be $\widehat h(\mathbf{y}) = \log \widehat g(\mathbf{x}) = \log c + \mathbf{a}^T\mathbf{y}$. Using Taylor's approximation of $h(\mathbf{y})$ around the point $\mathbf{y}_0$ in the feasible region and equating it with $\widehat h(\mathbf{y})$, it follows that
\begin{equation}
h(\mathbf{y}) \approx h(\mathbf{y}_0) + \nabla h(\mathbf{y}_0)^T(\mathbf{y} - \mathbf{y}_0) = \log c + \mathbf{a}^T\mathbf{y},
\label{eq:apprx1}
\end{equation}
for $\mathbf{y}\approx \mathbf{y_0}$. From \eqref{eq:apprx1}, we have $\mathbf{a} = \nabla h(\mathbf{y}_0)$, \textit{i.e.},
$$a_i = \left.\frac{x_i}{g(\mathbf{x})}\frac{\partial g}{\partial x_i}\right|_{\mathbf{x} = \mathbf{x}_0},$$
and\vspace*{-1mm}
$$c = \exp(h(\mathbf{y}_0) - \nabla h(\mathbf{y}_0)^T\mathbf{y}_0) = g(\mathbf{x}_0)\prod_{i=1}^5 x_{0,i}^{a_i},$$
where $x_{0,i}$ is an $i$th element of $\mathbf{x}_0$. 
We substitute the monomial approximation $\widehat g(x)$ of $g(x)$ in \eqref{g_mono} and use GP technique to solve \eqref{prob}. The aforementioned affine approximation is, however, imprecise if the optimal solution lies far from the initial guess $\mathbf{x}_0$ as the Taylor's approximation would be inaccurate. To overcome this problem, we take an iterative approach, where, if the current guess is $\mathbf{x}_k$, we obtain the Taylor's approximation about $\mathbf{x}_k$ and solve a GP again. Let the current solution of GP be $\mathbf{x}_{k+1}$. In the next iteration, we take Taylor's approximation around $\mathbf{x}_{k+1}$ and solve a GP again. We keep iterating in this fashion until the convergence. Since the problem \eqref{prob} is close to GP (as we have only one constraint in \eqref{prob} that is not a posynomial), the aforementioned iterative approach works well in our case and yields the optimal solution~\cite{boyd_gp}. If the obtained optimal solution contradicts with our initial assumption that $C_1^S - C_1^R \geq 0$ and $C_2^S - C_2^R \geq 0$, we move to other three cases discussed below.%\vspace*{-2mm}

\subsection*{\underline{Case II}: $C_1^S - C_1^R \geq 0$ and $C_2^S - C_2^R < 0$}
In this case, the secrecy rate is given by $C_S = (C_1^S - C_1^R)^+$, and we need to solve the problem \eqref{prob} with the following expressions for $f(\beta,\widetilde\beta,\gamma_1,\gamma_2, \gamma_J)$ and $g(\beta,\widetilde\beta,\gamma_1,\gamma_2, \gamma_J)$:
\begin{align*}
f(\beta,\widetilde\beta,\gamma_1,\gamma_2, \gamma_J) &=  [\widetilde\beta(\gamma_1 + \gamma_2 + \gamma_J) + 1]\\
&\times\lbrack 1 + (\gamma_1 + \gamma_2 + \gamma_J)(|h_2|^2\beta\eta + \widetilde\beta)\rbrack,  \\
g(\beta,\widetilde\beta,\gamma_1,\gamma_2, \gamma_J) &= (\widetilde\beta(\gamma_2 + \gamma_J) + 1)\\
\times\lbrack 1 + &(\gamma_1 + \gamma_2 + \gamma_J)(\widetilde\beta + |h_2|^2\beta\eta(\widetilde\beta\gamma_1 + 1))\rbrack.
\end{align*}
We again check if the assumption $C_1^S - C_1^R \geq 0$ and $C_2^S - C_2^R < 0$ is valid; if not, we move to the remaining two cases.

\subsection*{\underline{Case III}: $C_1^S - C_1^R < 0$ and $C_2^S - C_2^R \geq 0$} 
This case is similar to Case II, and only the subscripts 1 and 2 need to be interchanged in the expressions of $f(\beta,\widetilde\beta,\gamma_1,\gamma_2, \gamma_J)$ and $g(\beta,\widetilde\beta,\gamma_1,\gamma_2, \gamma_J)$. If the solution obtained does not satisfy the initial assumptions, we move to Case IV.

\subsection*{\underline{Case IV}: $C_1^S - C_1^R < 0$ and $C_2^S - C_2^R < 0$} 
In this case, the sum-secrecy rate is zero.

Algorithm~\ref{algo1} summarizes the aforementioned process of obtaining the optimal sum-secrecy rate and power allocation by solving \eqref{prob}.
\begin{algorithm}
\textbf{Input}  Total power $P$, Channel coefficients $h_1, h_2$, and $h_J$, Energy conversion efficiency $\eta$, Noise power $N_0$, Tolerance $\delta$\\
\textbf{Output} Power splitting ratio $\beta$, power $P_1$, $P_2$, and $P_J$, sum-secrecy rate $C_S$\\ 
\textbf{Initialize} $0 \leq P_{1,k}, P_{2,k}, P_{J,k} \leq P$, $0 < \beta_{k} < 1$ (Random initialization) with $k=0$ 
\begin{enumerate}
\item
\textbf{While} $|C_{S,k} - \ C_{S,k-1}| > \delta C_{S,k-1}$
\item
Find the monomial expression $\widehat{g}$ for $g$ using the Taylor's approximation around $\mathbf{x}_{k} = [\beta_k,\gamma_{1,k},\gamma_{2,k},\gamma_{J,k}]$
\item
$k = k+1$
\item
Solve \eqref{prob} with the monomial approximation $\widehat g$ to find $[\beta_k,\gamma_{1,k},\gamma_{2,k},\gamma_{J,k}]$
\item 
Assign $C_1^S, C_1^R,C_2^S$ and $C_2^R$ using above solution  
\item
\textbf{If} $C_1^S - C_1^R \geq 0$ and $C_2^S - C_2^R \geq 0$
\\
\quad \quad Go to step 1
\\
\textbf{Else}
\\
\quad\quad Proceed to Case II
\item
Check for Cases II, III, and IV in a similar fashion
\item Find the optimal $[\beta_k,\gamma_{1,k},\gamma_{2,k},\gamma_{J,k}]$ for the current iteration after going through all cases
\item
Assign $C_{S,k} = \frac{1}{2}\log \frac{g(\beta_k,\gamma_{1,k},\gamma_{2,k},\gamma_{J,k})}{f(\beta_k,\gamma_{1,k},\gamma_{2,k},\gamma_{J,k})}$
\item
\textbf{End While}\label{algo_1}
\end{enumerate}
 \caption{Solution of \eqref{prob} 
 }\label{algo1}
 \end{algorithm}

\section{Secure Communication with Imperfect CSI }
\label{sec:ICSI}
 
We now investigate the effect of imperfect CSI on sum-secrecy rate. We model the imperfection in channel knowledge as in \cite{csi_model}, where the channel coefficients are given as
\begin{equation}\label{imp_csi}
h_i = \hat h_i + \Delta h_i,
\end{equation}
for $i \in \lbrace 1,2, J\rbrace$. Here $\hat h_i$ is the estimated channel coefficient and $\Delta h_i$ is the error in estimation which is bounded as $|\Delta h_i| \leq \epsilon_i$. $\epsilon_i$ is the maximum possible error in estimating $h_i$ with respect to $S_1$ and $S_2$. We consider the worst case scenario where the relay knows all channel coefficients perfectly, while legitimate nodes $S_1$ and $S_2$ concede estimation errors according to \eqref{imp_csi}. In this case, SNRs at the relay corresponding to the messages $x_1$ and $x_2$ remain the same as in \eqref{snr_relay}.
The signal received at $S_2$ in the second slot is\vspace*{-2mm}

{{\small
\begin{align}
y_2 &= h_2\alpha\left(\!\sqrt{\widetilde \beta P_1}h_1x_1 + \sqrt{\widetilde \beta P_2}h_2x_2 + \sqrt{\widetilde \beta P_J}h_Jx_J + n_R \!\right) + n_2 \nonumber \\
&= (\hat h_2 + \Delta h_2)\alpha \bigg(\sqrt{\widetilde \beta P_1}(\hat h_1 + \Delta h_1)x_1 + \sqrt{\widetilde \beta P_2}(\hat h_2 + \Delta h_2)x_2 \nonumber \\ 
&+ \sqrt{\widetilde \beta P_J}(\hat h_J + \Delta h_J)x_J) + n_R\bigg) + n_2,
\end{align}}}\vspace*{-3mm}

\noindent where $\hat{h}_1$, $\hat{h}_2$, and $\hat{h}_J$ are the channel coefficients estimated by node $S_2$. Using these imperfect channel estimates, the node $S_2$ tries to cancel the self-interference and the known jammer's signal in the following manner:
\begin{align}
y_2 &= (\hat h_2 + \Delta h_2)\alpha(\sqrt{\widetilde \beta P_1}(\hat h_1 + \Delta h_1)x_1 \nonumber \\
&+ \sqrt{\widetilde \beta P_2}(\hat h_2 + \Delta h_2)x_2  
+ \sqrt{\widetilde \beta P_J}(\hat h_J + \Delta h_J)x_J) + n_R) \nonumber \\
 &+ n_2 - \underbrace{\hat h_2\alpha(\sqrt{\widetilde \beta P_2}\hat h_2x_2 + \sqrt{\widetilde \beta P_J}\hat h_Jx_J)}_{\text{imperfect~interference~cancellation}}. \label{eq:imp}
 \end{align}
It follows that 
 \begin{align}
y_2&= \hat h_2\alpha\sqrt{\widetilde \beta P_1}\hat h_1x_1 + (\hat h_2 + \Delta h_2)\alpha n_R + n_2 \nonumber \\
&+ \Delta h_2\alpha(\sqrt{\widetilde \beta P_1}\hat h_1x_1 + \sqrt{\widetilde \beta P_2}\hat h_2x_2 + \sqrt{\widetilde \beta P_J}\hat h_Jx_J) \nonumber \\
&+ \hat h_2\alpha(\sqrt{\widetilde \beta P_1}\Delta h_1x_1 + \sqrt{\widetilde \beta P_2}\Delta h_2x_2  
+ \sqrt{\widetilde \beta P_J}\Delta h_Jx_J).\nonumber
\end{align}\vspace*{-3mm}

%After substituting $\alpha = \sqrt{\beta\eta/\widetilde \beta}$,
\noindent As \eqref{eq:imp} shows, due to the imperfect CSI, $S_2$ cannot cancel the jamming signal and the self-interference completely. Here we ignore the smaller terms of the form $\Delta h_i \Delta h_j$ as they will be negligible compared to other terms. The received SNR at $S_2$ is thus given by \eqref{snr_s21} at the top of the next page.
\begin{figure*}
\begin{equation}\label{snr_s21}
\text{SNR}_{S_2} \!=\! \frac{|\hat h_2|^2\alpha^2\widetilde\beta P_1|\hat h_1|^2}{N_0(|\hat h_2 + \Delta h_2|^2\alpha^2 + 1) + \alpha^2\widetilde \beta(|\Delta h_2|^2(P_1|\hat h_1|^2 + P_2|\hat h_2|^2 + P_J|\hat h_J|^2) + |\hat h_2|^2(P_1|\Delta h_1|^2 + P_2|\Delta h_2|^2 + P_J|\Delta h_J|^2))}.
\end{equation}\vspace*{-2mm}
\end{figure*}
Using the triangle inequality, it follows that
$$|\hat h_i| - |\Delta h_i| \leq h_i \leq |\hat h_i| + |\Delta h_i|,~~\forall~ i \in \lbrace 1,2,J \rbrace.\vspace*{-1mm} $$
The worst case secrecy rate will occur when
$$h_i = |\hat h_i| + |\Delta h_i| = |\hat h_i| + \epsilon_i,~~\forall~ i \in \lbrace 1,2,J \rbrace, $$ 
and this will happen when the phase of $h_i$ and $\Delta h_i$ are the same and $\Delta h_i$ concedes maximum error, \textit{i.e.}, $|\Delta h_i| = \epsilon_i$. Then the worst case SNR (denoted by SNR$_{S_2}^{wc}$) at node $S_2$ is given by \eqref{snr_s2} at the top of the next page. 
\begin{figure*}
\begin{equation}\label{snr_s2}
\text{SNR}_{S_2}^{wc} = \frac{|\hat h_2|^2\alpha^2\widetilde\beta P_1|\hat h_1|^2}{N_0((|\hat h_2| + \epsilon_2)^2\alpha^2 + 1) + \alpha^2\widetilde \beta\epsilon_2^2(P_1|\hat h_1|^2 + P_2|\hat h_2|^2 + P_J|\hat h_J|^2) + \alpha^2\widetilde\beta|\hat h_2|^2(P_1\epsilon_1^2 + P_2\epsilon_2^2 + P_J\epsilon_J^2)}.
\end{equation}\vspace*{-3mm}
\end{figure*}
Similarly the worst case SNR (denoted by SNR$_{S_1}^{wc}$) at $S_1$ is given by \eqref{snr_s1} at the top of the next page. In \eqref{snr_s1}, we again denote estimated channels by $\hat h_1$, $\hat h_2$, and $\hat h_J$ for brevity, but these values may be different from those estimated by $S_2$.
\begin{figure*}
\begin{equation}\label{snr_s1}
\text{SNR}_{S_1}^{wc} = \frac{|\hat h_1|^2\alpha^2\widetilde\beta P_2|\hat h_2|^2}{N_0((|\hat h_1| + \epsilon_1)^2\alpha^2 + 1) + \alpha^2\widetilde \beta\epsilon_1^2(P_1|\hat h_1|^2 + P_2|\hat h_2|^2 + P_J|\hat h_J|^2) + \alpha^2\widetilde\beta|\hat h_1|^2(P_1\epsilon_1^2 + P_2\epsilon_2^2 + P_J\epsilon_J^2)}.
\end{equation}
\hrulefill\vspace*{-4mm}
\end{figure*}

%Since we aim to calculate the worst case secrecy rate, we would like to minimize the information throughput for $S_1$, while at the same time maximizing throughput for relay. From \eqref{snr_r1_r2}, we can calculate an upper bound on SNR$_{R_1}$ and SNR$_{R_2}$ as
%\begin{eqnarray*}
%\text{SNR}_{R_1} \leq \frac{\widetilde\beta P_2|\hat h_2|^2}{\widetilde\beta (P_1(|\hat h_1| - \epsilon)^2 + P_J(|\hat h_J| - \epsilon)^2) + N_0}\\
%\text{SNR}_{R_2} \leq \frac{\widetilde\beta P_1|\hat h_1|^2}{\widetilde\beta (P_2(|\hat h_2| - \epsilon)^2 + P_J(|\hat h_J| - \epsilon)^2) + N_0}.
%\end{eqnarray*}
%Note that these bounds cannot be satisfied simultaneously. Nevertheless, they allow us to find a lower bound on the achievable secrecy rate.
%, which is not very straightforward. Hence, we use scientific computing tools like MATLAB to find the best case throughput scenario for the relay. 
Using these worst case SNRs, we maximize the worst case sum-secrecy rate and solve for the corresponding optimal power allocation and $\beta$ using SGP as done for problem in \eqref{prob}, \textit{i.e.}, for the case of perfect CSI.

\section{Numerical Results and Discussions}\label{numerical_results}

\subsection{Effect of Power Splitting Ratio $\beta$} 

\begin{figure}
\centering
\includegraphics[scale=0.6]{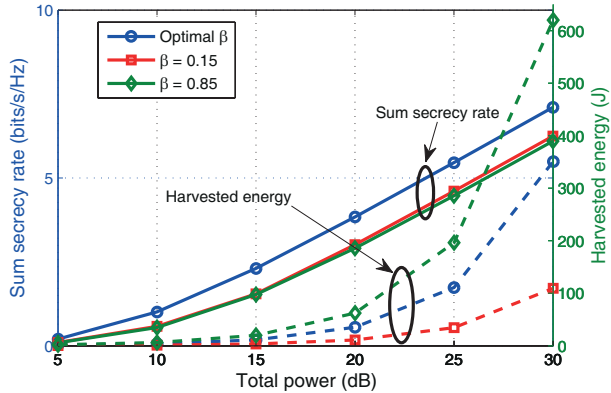}\vspace*{-2mm}
\caption{Effect of $\beta$ on harvested energy at relay and the sum-secrecy rate.}
\label{fig_ph}\vspace*{-4mm}
\end{figure}

\begin{figure}
\centering
\includegraphics[scale=0.6]{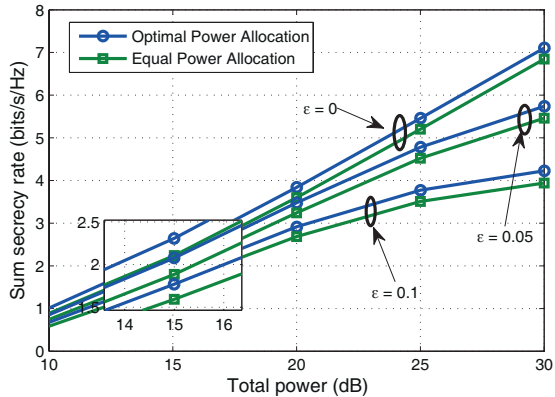}\vspace*{-2mm}
\caption{Effect of power allocation on sum-secrecy rate.}
\label{fig_power_alloc}\vspace*{-4mm}
\end{figure}

Fig.~\ref{fig_ph} shows the sum-secrecy rate (left y-axis) and the harvested energy (right y-axis) versus the total power budget for a random channel realization: $|h_1|^2 = 0.6647, |h_2|^2 = 2.9152$, and $|h_J|^2 = 1.3289$. We set $\eta = 0.5$ and $N_0 = 1$. Higher $\beta$ (= 0.85) than the optimal $\beta$ (the solution of the problem \eqref{prob}) results in higher harvested energy, which increases relay's transmit power, but the reduced strength of the received information signal at the relay (thus at nodes $S_1$ and $S_2$) due to higher $\beta$ dominates the secrecy performance of the system. A lower $\beta$ (= 0.15) ensures more power for the information processing at relay, but this reduces the harvested energy (reducing its transmit power to forward the information) and increases the chances of relay eavesdropping the secret message. As a result, the sum-secrecy rate reduces.

\subsection{Effect of Power Allocation}
For different values of maximum channel estimation errors, Fig.~\ref{fig_power_alloc} compares the sum-secrecy rate when the total power is allocated optimally (obtained by solving the problem \eqref{prob}) and equally among nodes $S_1$, $S_2$, and jammer $J$ for the same system parameters used to obtain Fig.~\ref{fig_ph}. For exposition, we consider $\epsilon_1 = \epsilon_2 = \epsilon_J = \epsilon$ in numerical results. The case $\epsilon = 0$ corresponds to the perfect CSI at $S_1$ and $S_2$. Since the equal power allocation does not use channel conditions optimally, it suffers a loss in sum-secrecy rate as expected. Due to the error in channel estimation, the nodes $S_1$ and $S_2$ cannot cancel the self-interference (information signals sent to the relay in the first slot) and the jamming signal perfectly from the received signal in the second slot. This reduces the SNR at legitimate nodes $S_1$ and $S_2$, which further reduces the sum-secrecy rate. 

%More allocation of power to jammer would surely degrade the decoding capacity of the relay, but it also has two important negative side effects. First, the jammer is being allocated power from the limited power bank which otherwise could be used by $S_1$ and $S_2$ to improve information throughput. Second, the relay forwards the unintended jammer signal to both $S_1$ and $S_2$ in the second hop using the harvested energy $P_H$. More harvested energy would be wasted in forwarding the jammer signal if the power allocated to jammer is more. This results in an interesting trade-off between jammer power and $JR$ distance as shown in Fig. \ref{pj}. Here, the distances $S_1R$ and $S_2R $ are five meters each and total power $P=30$ dB. One can also notice that secrecy rate decreases with increasing jammer to relay distance.  

\begin{figure}
\centering
\includegraphics[scale=0.6]{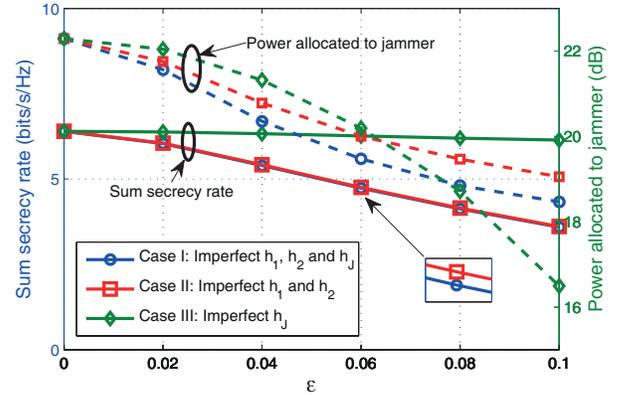}\vspace*{-2mm}
\caption{Effect of $\epsilon$ on sum-secrecy rate and and power $P_J$ allocated to the jammer, $|h_1^2| = 1.2479$, $|h_2|^2 = 1.4484$, and $|h_J|^2 = 6.0162$, $P = 30~\mathrm{dB}.$}
\label{fig_ep_pj}\vspace*{-4mm}
\end{figure}

\subsection{Effect of Imperfect CSI}
Fig.~\ref{fig_ep_pj} shows three cases based on the knowledge of channel conditions at $S_1$ and $S_2$.\footnote{These three cases in Fig.~\ref{fig_ep_pj} should not be confused with four cases considered in Section~\ref{prob_formulation}.}
% The blue curve shows the variation of secrecy rate (solid line) and optimal power allocated to jammer (dotted line) when all the channels $h_1, h_2$ and $h_J$ are imperfectly known at $S_1$ and $S_2$. The red curve symbolize the case when $h_1$ and $h_2$ are imperfectly known and $h_j$ is perfectly estimated and the blue curve represents a vice-versa scenario. 
The sum-secrecy rate in Case II is slightly better than that in Case I, because in Case II, a higher fraction of the total power is allocated to the jammer (see the right y-axis of Fig.~\ref{fig_ep_pj}) to use the perfect channel knowledge about $h_J$. But this has a side-effect: the imperfect CSI about $h_1$ and $h_2$ leads to higher interference from the jammer to $S_1$ and $S_2$. As a result, Case II does not gain much compared to Case I in terms of the sum-secrecy rate. Under Case III, the sum-secrecy rate is the highest, because $S_1$ and $S_2$ can cancel the jamming signal more effectively as they have imperfect CSI about only one channel. When $\epsilon$ is small enough (less than 0.06 in this case), the power allocated to the jammer in Case III is higher than that in Cases I and II. This is because when $\epsilon$ is small, if we allocate the power to $S_1$ and $S_2$ instead of jammer, it increases relay's chances of eavesdropping the information due to the increased received power, which dominates the detrimental effect incurred due to imperfect cancellation of jammer's signal at $S_1$ and $S_2$. But if $\epsilon$ goes beyond a threshold, the loss in the secrecy rate due to the imperfect cancellation of jammer's interference dominates, and the system is better off by allocating more power to $S_1$ and $S_2$ and using each other's signals to confuse the relay. Hence the power allocated to jammer in Case III is smaller than that in Cases I and II at higher $\epsilon$. In Case III, the redistribution of the power from jammer to $S_1$ and $S_2$ with the increase in $\epsilon$ keeps the sum-secrecy rate almost the same.

\section{Concluding Remarks and Future Directions}
In a two-way untrusted relay scenario, though the signal from one source can indirectly serve as an artificial noise to the relay while processing other source's signal, the non-zero power allocated to the jammer implies that the assistance from an external jammer can still be useful to achieve a better secrecy rate. But the knowledge of two sources about channel conditions decides the contribution of the jammer in achieving the secure communication. For example, as the channel estimation error on any of the channel increases, the power allocated to the jammer decreases to subside the interference caused at the sources due to the imperfect cancellation of the jamming signal. The optimal power splitting factor balances between the energy harvesting and the information processing at relay. Hence the joint allocation of the total power and the selection of the power splitting factor are necessary to maximize the sum-secrecy rate. 

\textit{Future directions}: There are several interesting future directions that are worth investigating. First the proposed model can be extended to general setups such as multiple antennas at nodes and multiple relays. Another interesting future direction is to investigate the effect of the placement of the jammer and the relay, which also incorporates the effect of path loss. Third we have considered the bounded uncertainty model to characterize the imperfect CSI. Extension to other models of imperfect CSI such as the model where only channel statistics are known is also possible.

\bibliographystyle{IEEEtran}
\bibliography{IEEEabrv,bibli}

% Generated by IEEEtran.bst, version: 1.14 (2015/08/26)
\begin{thebibliography}{10}
\providecommand{\url}[1]{#1}
\csname url@samestyle\endcsname
\providecommand{\newblock}{\relax}
\providecommand{\bibinfo}[2]{#2}
\providecommand{\BIBentrySTDinterwordspacing}{\spaceskip=0pt\relax}
\providecommand{\BIBentryALTinterwordstretchfactor}{4}
\providecommand{\BIBentryALTinterwordspacing}{\spaceskip=\fontdimen2\font plus
\BIBentryALTinterwordstretchfactor\fontdimen3\font minus
  \fontdimen4\font\relax}
\providecommand{\BIBforeignlanguage}[2]{{%
\expandafter\ifx\csname l@#1\endcsname\relax
\typeout{** WARNING: IEEEtran.bst: No hyphenation pattern has been}%
\typeout{** loaded for the language `#1'. Using the pattern for}%
\typeout{** the default language instead.}%
\else
\language=\csname l@#1\endcsname
\fi
#2}}
\providecommand{\BIBdecl}{\relax}
\BIBdecl

\bibitem{relay2006}
B.~Rankov and A.~Wittneben, ``Achievable rate regions for the two-way relay
  channel,'' in \emph{Proc. 2006 IEEE ISIT}, pp. 1668--1672.

\bibitem{yener_two_way}
M.~Chen and A.~Yener, ``Multiuser two-way relaying: detection and interference
  management strategies,'' \emph{IEEE Trans. Wireless Commun.}, vol.~8, no.~8,
  pp. 4296--4305, Aug. 2009.

\bibitem{lav}
L.~Varshney, ``Transporting information and energy simultaneously,'' in
  \emph{Proc. 2008 IEEE ISIT}, pp. 1612--1616.

\bibitem{rui4}
X.~Zhou, R.~Zhang, and C.~K. Ho, ``Wireless information and power transfer:
  Architecture design and rate-energy tradeoff,'' \emph{IEEE Trans. Commun.},
  vol.~61, no.~11, pp. 4754--4767, Nov. 2013.

\bibitem{nasir}
A.~A. Nasir, X.~Zhou, S.~Durrani, and R.~A. Kennedy, ``Relaying protocols for
  wireless energy harvesting and information processing,'' \emph{IEEE Trans.
  Wireless Commun.}, vol.~12, no.~7, pp. 3622--3636, July 2013.

\bibitem{chen1}
H.~Chen, Y.~Jiang, Y.~Li, Y.~Ma, and B.~Vucetic, ``A game-theoretical model for
  wireless information and power transfer in relay interference channels,'' in
  \emph{Proc. 2014 IEEE ISIT}, pp. 1161--1165.

\bibitem{sanket3}
S.~S. Kalamkar and A.~Banerjee, ``Interference-aided energy harvesting:
  {Cognitive} relaying with multiple primary transceivers,'' \emph{IEEE Trans.
  Cogn. Commun. Netw.}, accepted for publication.

\bibitem{liu1}
Z.~Chen, B.~Xia, and H.~Liu, ``Wireless information and power transfer in
  two-way amplify-and-forward relaying channels,'' in \emph{Proc. 2014 IEEE
  GLOBALSIP}, pp. 168--172.

\bibitem{duong_two_way}
Y.~Liu, L.~Wang, M.~Elkashlan, T.~Q. Duong, and A.~Nallanathan, ``Two-way
  relaying networks with wireless power transfer: Policies design and
  throughput analysis,'' in \emph{Proc. 2014 IEEE GLOBECOM}, pp. 4030--4035.

\bibitem{wyner}
A.~D. Wyner, ``The wire-tap channel,'' \emph{Bell Syst. Tech. J.}, vol.~54,
  no.~8, pp. 1355--1387, Jul. 1975.

\bibitem{quan}
Q.~Li, Q.~Zhang, and J.~Qin, ``Secure relay beamforming for simultaneous
  wireless information and power transfer in nonregenerative relay networks,''
  \emph{IEEE Trans. Veh. Technol.}, vol.~63, no.~5, pp. 2462--2467, June 2014.

\bibitem{xing2}
H.~Xing, Z.~Chu, Z.~Ding, and A.~Nallanathan, ``Harvest-and-jam: Improving
  security for wireless energy harvesting cooperative networks,'' in
  \emph{Proc. 2014 IEEE GLOBECOM}, pp. 3145--3150.

\bibitem{chen2016}
X.~Chen, J.~Chen, H.~Zhang, Y.~Zhang, and C.~Yuen, ``On secrecy performance of
  a multi-antenna jammer aided secure communications with imperfect {CSI},''
  \emph{IEEE Trans. Veh. Technol.}, vol.~65, no.~10, pp. 8014--8024, Oct. 2016.

\bibitem{he}
X.~He and A.~Yener, ``Cooperation with an untrusted relay: A secrecy
  perspective,'' \emph{IEEE Trans. Inf. Theory}, vol.~56, no.~8, pp.
  3807--3827, Aug. 2010.

\bibitem{huang}
J.~Huang, A.~Mukherjee, and A.~L. Swindlehurst, ``Secure communication via an
  untrusted non-regenerative relay in fading channels,'' \emph{IEEE Trans.
  Signal Process.}, vol.~61, no.~10, pp. 2536--2550, May 2013.

\bibitem{wang}
L.~Wang, M.~Elkashlan, J.~Huang, N.~H. Tran, and T.~Q. Duong, ``Secure
  transmission with optimal power allocation in untrusted relay networks,''
  \emph{IEEE Wireless Commun. Lett.}, vol.~3, no.~3, pp. 289--292, June 2014.

\bibitem{li2}
L.~Sun, P.~Ren, Q.~Du, Y.~Wang, and Z.~Gao, ``Security-aware relaying scheme
  for cooperative networks with untrusted relay nodes,'' \emph{IEEE Commun.
  Lett.}, vol.~19, no.~3, pp. 463--466, Mar. 2015.

\bibitem{park2015}
K.-H. Park and M.-S. Alouini, ``Secure amplify-and-forward untrusted relaying
  networks using cooperative jamming and zero-forcing cancelation,'' in
  \emph{Proc. 2015 IEEE PIMRC}, pp. 234--238.

\bibitem{zhang2012}
R.~Zhang, L.~Song, Z.~Han, and B.~Jiao, ``Physical layer security for two-way
  untrusted relaying with friendly jammers,'' \emph{IEEE Trans. Veh. Technol.},
  vol.~61, no.~8, pp. 3693--3704, Oct. 2012.

\bibitem{wang2016}
D.~Wang, B.~Bai, W.~Chen, and Z.~Han, ``Secure green communication via
  untrusted two-way relaying: A physical layer approach,'' \emph{IEEE Trans.
  Commun.}, vol.~64, no.~5, pp. 1861--1874, May 2016.

\bibitem{sanket}
S.~S. Kalamkar and A.~Banerjee, ``Secure communication via a wireless energy
  harvesting untrusted relay,'' \emph{IEEE Trans. Vech. Technol.}, vol.~66,
  no.~3, pp. 2199--2213, Mar. 2017.

\bibitem{meng}
M.~Zhao, S.~Feng, X.~Wang, M.~Zhang, Y.~Liu, and H.~Fu, ``Joint power splitting
  and secure beamforming design in the wireless-powered untrusted relay
  networks,'' in \emph{Proc. 2015 IEEE GLOBECOM}, pp. 1--6.

\bibitem{mousa}
D.~J. Su, S.~A. Mousavifar, and C.~Leung, ``Secrecy capacity and wireless
  energy harvesting in amplify-and-forward relay networks,'' in \emph{Proc.
  2015 IEEE PACRIM}, pp. 258--262.

\bibitem{boyd_gp}
S.~Boyd, S.-J. Kim, L.~Vandenberghe, and A.~Hassibi,
  ``\BIBforeignlanguage{English}{A tutorial on geometric programming},''
  \emph{\BIBforeignlanguage{English}{Optimization and Engineering}}, vol.~8,
  no.~1, pp. 67--127, 2007.

\bibitem{hayashi}
M.~Hayashi and R.~Matsumoto, ``Construction of wiretap codes from ordinary
  channel codes,'' in \emph{Proc. 2010 IEEE ISIT}, pp. 2538--2542.

\bibitem{tekin}
E.~Tekin and A.~Yener, ``The general {Gaussian} multiple-access and two-way
  wiretap channels: Achievable rates and cooperative jamming,'' \emph{IEEE
  Trans. Inf. Theory}, vol.~54, no.~6, pp. 2735--2751, June 2008.

\bibitem{csi_model}
Z.~Xiang and M.~Tao, ``Robust beamforming for wireless information and power
  transmission,'' \emph{IEEE Wireless Commun. Lett.}, vol.~1, no.~4, pp.
  372--375, Aug. 2012.

\end{thebibliography}
%\bibliography{IEEEabrv,References_ICCw}
\end{document}